\documentclass[twocolumn,nofootinbib,amsmath,amssymb,floatfix,usenatbib,prd,useAMS]{revtex4}
\usepackage{graphicx,epsfig}
\usepackage{bm}
\usepackage{amsmath}
\usepackage{epstopdf}
\def\fdg{\hbox{$.\!\!^\circ$}}
\renewcommand{\l}{\ell}
\newcommand\eq[1]{Eq.~(\ref{#1})}
\newcommand\eqs[2]{Eqs.~(\ref{#1}) and (\ref{#2})}

\newcommand\slabel[1]{\label{#1}}
\newcommand{\Mp}{M_{\rm  pl}}
\newcommand{\alm}{a_{\ell m}}

\newcommand{\beq}{\begin{equation}}

\newcommand{\eeq}{\end{equation}}
\newcommand{\beqa}{\begin{eqnarray}}
\newcommand{\eeqa}{\end{eqnarray}}

\newcommand{\wj}[6]{\left(
                           \begin{array}{ccc}
        \! #1\! & #2\!  & #3\!  \\
        \! #4\! & #5\!  & #6\!
                           \end{array}
                   \right)}

\def\simlt{\lesssim}
\def\simgt{\gtrsim}

\newcommand{\apjs}{ApJS}
\newcommand{\mnras}{MNRAS}
\newcommand{\physrev}{Phys. Rev.}
\newcommand{\physrevlett}{Phys. Rev. Lett.}

\begin{document}

\title{Spontaneous Isotropy Breaking: A Mechanism for CMB Multipole Alignments}
\author{Christopher Gordon, Wayne Hu, Dragan Huterer and Tom Crawford}
\affiliation{
Kavli Institute for Cosmological Physics, Department of Astronomy and Astrophysics,
and Enrico Fermi Institute,  University of Chicago, Chicago IL 60637 
}

\begin{abstract}
\baselineskip 11pt We introduce a class of models in which statistical isotropy is 
broken spontaneously in the CMB by a non-linear response to long-wavelength 
fluctuations in a  mediating field.  These fluctuations appear as 
a gradient locally and pick out a single 
preferred direction. The non-linear response 
imprints this direction in a range of multipole  moments. We consider two 
manifestations of isotropy breaking: additive contributions and  multiplicative 
modulation of the intrinsic anisotropy.  Since WMAP exhibits an alignment of power  
deficits, an additive contribution is {\it less} likely to produce the observed alignments 
than the  usual isotropic fluctuations, a fact which we illustrate with an explicit 
cosmological model of  long-wavelength quintessence fluctuations.  This problem 
applies to other models involving  foregrounds or background anisotropy that seek to 
restore power to the CMB.  Additive models  that account directly for the observed 
power exacerbate the low power of the intrinsic fluctuations.  Multiplicative models can 
overcome these  difficulties.  We construct a proof of principle model that significantly
improves the likelihood and generates stronger alignments than WMAP 
in 30-45\% of realizations.
\end{abstract}
\maketitle

\section{Introduction}

Data from the first year of WMAP has largely confirmed the standard
cosmological model based on nearly scale-invariant statistically homogeneous
and isotropic fluctuations \cite{Bennett2003}.  Yet interestingly the large
angle deficit in temperature fluctuations versus the standard model, first seen
by COBE \cite{Smoetal92}, was both confirmed and sharpened.  This result is
unlikely at between 0.7\% and 10\%, depending on the assumptions, the choice of
statistic, and the analysis
\citep{Efstathiou2003,Slosar2004,Bielewicz2004,O'Dwyer2004}.  Cosmological
explanations for a deficit of power involve breaking scale-invariance
initially, dynamically, or topologically.

More recently, however, several other anomalies have been noted that point to a
possible breaking of statistical isotropy on large angles.  Statistically
isotropy requires that all of the multipole moments of the CMB temperature
field be uncorrelated at the two point level.  Yet the WMAP data exhibit
curious correlations or alignments in the temperature field.  In particular,
the octopole of the CMB is highly planar and aligned with the quadrupole
\citep{TOH,deOliveira2004}.  Specifically the normalized angular momentum of
the octopole is anomalously large at the 1/20 level and the angular momentum
directions of the quadrupole and octopole are anomalously close at the 1/60
level.  These and other alignments in the large-angle CMB have been further
discussed and analyzed by Refs.\ \citep{Copi2004,Schwarz2004,Land2005a,
Land2005b,Bielewicz2004, SS2004} and comprehensively studied in Ref.\
\citep{Copi2005}.

While the {\it a posteriori} nature of these statistics makes a strict
interpretation of these probabilities difficult, the anomalies are
clearly significant enough to merit serious attention.  There are at least
three classes of physical explanations: systematic, astrophysical, and
cosmological.  The WMAP instrument and time stream \cite{Hinetal03} have passed
stringent tests that would make a systematic error of this magnitude and type
difficult to conceive.  An astrophysical explanation involving foreground contamination
is certainly possible.  However known foregrounds are accounted for very
successfully (e.g.\ \citep{Bennett_foregr, Finkbeiner2003}) and the
contamination would need to be of order the intrinsic CMB temperature
anisotropy to account for the full effect.  Still foreground and systematic
contamination may weaken the significance of at least some of the statistical
anomalies typically at the expense of others \cite{SS2004,Bielewicz2005,Copi2005}.

The final possibility is that the anomalies have a cosmological explanation.
Large-scale homogeneity and isotropy are clearly good assumptions for the
universe as a whole.  In this study, we examine the possibility that
statistical isotropy is spontaneously broken in the fluctuations observed from
a given spatial position and not in the fundamental theory.  Correspondingly we
will take the all-sky WMAP data \cite{TOH} 
at face value and ignore the possibility of
foreground and systematic contamination.  We begin in \S \ref{sec:spontaneous}
with the general mechanism and proceed in \S \ref{sec:additive} and \S
\ref{sec:multiplicative} to describe two different implementations of the
mechanism.  We discuss these results in \S \ref{sec:discussion}.

\section{Spontaneous Isotropy Breaking}
\label{sec:spontaneous}

We introduce a general mechanism for the breaking of isotropy in large angle
fluctuations in \S\ref{sec:mechanism}.  Under this mechanism statistical
isotropy is preserved in the full theory but is spontaneously broken due to
long-wavelength field fluctuations that appear as a gradient locally to the
observer.  A non-linear response to the field by CMB temperature fluctuations
carries this preferred direction into a spectrum of multipoles.  In
\S\ref{sec:statistical} we discuss the statistical treatment of the induced
correlations between multipole moments used in the following sections.

\subsection{Non-linear Gradient Modulation}
\label{sec:mechanism}

There are two ingredients in our mechanism for breaking the statistical
isotropy.  First, we require the presence of a field whose spatial fluctuations are
dominated by long-wavelength contributions.  Locally, around an observer, the
structure in the field looks like a pure gradient.  Given that a gradient must
pick out a single direction, its presence provides the preferred axis for the
breaking of isotropy.  Because the long wavelength field is itself a random
fluctuation, the underlying full theory retains statistical homogeneity and
isotropy.  Statistical isotropy is only apparently broken locally, and we refer
to this as the spontaneous isotropy breaking.

The second ingredient is a non-linear response of the CMB temperature
perturbations to this field.  If the response were purely linear, then the
field gradient would produce a dipole in the CMB.  If the non-linearity were
small, the higher order anisotropy would be increasingly suppressed with
multipole number.  A familiar example of this latter case is the Doppler effect 
due to the motion of the solar system through the CMB.
The observed dipole is of course created by this motion.   The effect on the
quadrupole is a small correction to the true quadrupole proportional to $v^2$,
while the corrections to the higher multipoles are entirely negligible
\cite{Smoetal92}.  Here we seek to break isotropy in at least the quadrupole
and octopole.  Therefore the non-linearity must be substantial.

Let us now parameterize a class of model that possesses these ingredients.  A
pure gradient projected across the sky along some axis $\hat {\bf x}_3$ has the
structure
\begin{equation}
G(\hat{\bf n}) \propto (1 +  G_1 \hat{\bf n}\cdot \hat{\bf x}_3) = (1 + G_1 \cos\theta)\,,
\end{equation} 
where $\hat{\bf n}$ is the sky direction, $G_{1}$ is the relative amplitude of
the gradient and $\theta$ is the polar angle from the preferred axis.  A
non-linear response to this gradient can be described as a general function of
$G(\hat{\bf n})$, $F[ G(\hat{\bf n}) ]$.  This function can in turn be represented by its
spherical harmonic coefficients with the pole aligned in the direction of the
gradient axis
\begin{equation}
F[ G(\hat{\bf n}) ] \equiv w(\hat{\bf n}) 
= \sum_{\ell} w_\ell Y_{\ell 0}(\hat{\bf n}) \,.
\slabel{eqn:wdecomp}
\end{equation}
The azimuthal symmetry around the preferred axis of a gradient limits the
coefficients to $m=0$.

In general, the CMB may exhibit a non-linear response in two different ways.
It may produce an additive effect that is uncorrelated with the intrinsic
anisotropy or it may produce a multiplicative modulation of this anisotropy.
Furthermore, since the intrinsic anisotropy arises from a multitude of physical
effects, it is possible that only one of them carries a non-linear response.

Thus let us model the temperature fluctuation field of the CMB as
\begin{eqnarray}
T(\hat{\bf n}) \equiv A(\hat{\bf n}) + f  [1+ w(\hat{\bf n}) ] B(\hat{\bf n}) \,,
\label{eqn:generalform}
\end{eqnarray}
where $A(\hat{\bf n})$ and $B(\hat{\bf n})$ are statistically isotropic
Gaussian random fields.   In the trivial
case where $B(\hat{\bf n})=1$, the non-linear response is additive with an
amplitude proportional to the parameter $f$.  Alternately $A$ and $B$ may
represent two different physical sources of the intrinsic anisotropy, e.g. the
Sachs-Wolfe effect and the integrated Sachs-Wolfe effect.  The non-linear
modulation then produces a multiplicative breaking of isotropy in one of the
effects.

\subsection{Statistical Framework}
\label{sec:statistical}

It is convenient to describe the statistical properties of the model
in terms of the multipole moments of the fields.
The observed temperature field of \eq{eqn:generalform} 
becomes
\begin{eqnarray}
T(\hat{\bf n}) &=& \sum_{\ell m} t_{\ell m} Y_{\ell m}(\hat{\bf n}) \,, 
\end{eqnarray}
and similarly for the underlying isotropic fields
\begin{eqnarray}
A(\hat{\bf n}) &=& \sum_{\ell m} a_{\ell m} Y_{\ell m}(\hat{\bf n})\,,  \nonumber\\
B(\hat{\bf n}) &=& \sum_{\ell m} b_{\ell m} Y_{\ell m}(\hat{\bf n}) \,. \slabel{eqn:harmonicdecomp}
\end{eqnarray}
The assumption of statistical isotropy for the underlying fields $A$ and $B$
requires that their covariance matrices satisfy
\begin{eqnarray}
\langle  a_{\ell m}^* a_{\ell' m' }\rangle &=& \delta_{\ell \ell'} \delta_{m m'} C_\ell^{aa} \,,
\nonumber\\
\langle  a_{\ell m}^* b_{\ell' m' }\rangle &=& \delta_{\ell \ell'} \delta_{m m'} C_\ell^{ab} \,, 
\nonumber\\
\langle  b_{\ell m}^* b_{\ell' m' }\rangle &=& \delta_{\ell \ell'} \delta_{m m'} C_\ell^{bb} \,.
\end{eqnarray}
However statistical isotropy is not preserved in the observed
temperature field $T(\hat{\bf n})$.

Taking the multipole moments of \eq{eqn:generalform}, the product $w(\hat{\bf
n})B(\hat{\bf n})$ becomes a convolution
\begin{equation}
t_{\ell m}  = a_{\ell m} + f b_{\ell m} +  f \sum_{\ell_1 \ell_2}
R_{\ell m}^{\ell_1 \ell_2} b_{\ell_2 m}
\label{eqn:tlm}
\end{equation}
with a coupling matrix written in terms of Wigner 3j symbols
\begin{eqnarray}
R_{\ell m}^{\ell_1 \ell_2} &\equiv & 
(-1)^m \sqrt{ (2\ell+1)(2\ell_1+1)(2\ell_2+1) \over 4\pi} \nonumber\\
&&\times \wj{\ell_1}{\ell_2}{\ell}{0}{0}{0} 
\wj{\ell_1}{\ell_2}{\ell}{0}{m}{-m} w_{\ell_1}\,.
\end{eqnarray}
The ensemble average of the multipole moments becomes
\begin{eqnarray}
\langle  t_{\ell m}^* t_{\ell' m }\rangle  &=& 
\delta_{\ell \ell'} [C_\ell^{aa} + 2 f C_\ell^{ab} + f^2 C_\ell^{bb}]  \nonumber\\
&&+
f\sum_{\ell_1} \left[ R_{\ell' m}^{\ell_1 \ell} (C_{\ell}^{ab}+f C_{\ell}^{bb}) + 
  (\ell \leftrightarrow \ell')\right]
\nonumber\\
&&
+f^2 \sum_{\ell_1 \ell_1' \ell_2} R_{\ell m}^{\ell_1 \ell_2} R_{\ell' m}^{\ell_1' \ell_2} 
C_{\ell_2}^{bb}\,. \slabel{eqn:covmatrix}
\end{eqnarray}
Here the ensemble average is over realizations of the fields $A$ and $B$ with
a fixed field $w$.
The field $T({\bf n})$ exhibits broken  statistical isotropy with
 different variances in $m$ modes of a given $\ell$ and a covariance
 between different $\ell$ modes.

\begin{figure}
\epsfig{file=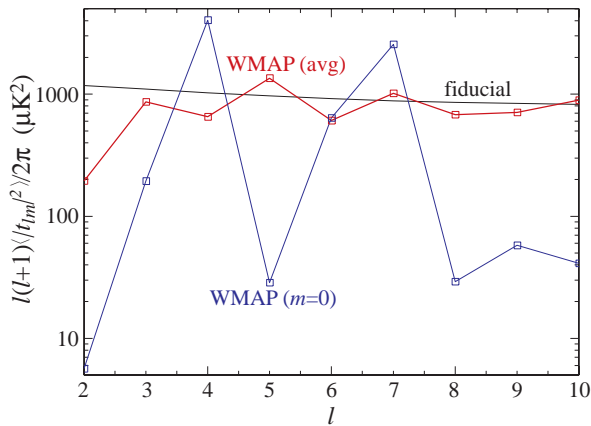, width=3.25in}
\caption{\footnotesize The fiducial cosmological model and the WMAP power in
$\ell$ averaged over $m$ (``avg'') and from azimuthally symmetric (``$m=0$")
components in the dipole frame.  In addition to an overall deficit of power in
the quadrupole, the WMAP quadrupole and octopole exhibit a further reduction of
power along the dipole direction. The fiducial cosmological model is defined by
$\Omega_{m}=0.27$, $h=0.72$, $\Omega_{b}h^{2}=0.024$, $\delta_{\zeta}= 5.07
\times 10^{-5}$ and $\tau=0.17$.  }
\label{fig:wmap}
\end{figure}

For the simplest case of an additive effect $B(\hat{\bf n})=1$ and is not a
 stochastic field.  We can however consider its effective power spectrum to be
 $C_\ell^{bb} =4\pi\delta_{\ell 0}$.  The ensemble average of
 \eq{eqn:covmatrix} reduces to
\begin{eqnarray}
\langle  t_{\ell m}^* t_{\ell' m }\rangle  &=& 
\delta_{\ell \ell'}  C_\ell^{aa} + f^2 w_\ell w_{\ell'} \delta_{m0}  \,,
\end{eqnarray}
where $f w_\ell$ is a deterministic contribution or mean to the
Gaussian distribution of $t_{\ell 0}$.  A
multiplicative model manifests more complex phenomenology.  Here each
$w_{\ell}$ correlates the observed multipoles across a range $\Delta \ell =
\ell$.  With a spectrum of $w_{\ell}$ the maximum range of the correlation is
determined by the highest non-negligible mode in the spectrum. We will explore
the additive and multiplicative versions in \S \ref{sec:additive} and \S
\ref{sec:multiplicative} respectively.

A hallmark of this mechanism is that the modulation is azimuthally symmetric and
hence couples harmonics with the same $m$ in the
preferred frame. Note however that in a general coordinate system, for example
galactic coordinates, different $m$ values will also be correlated.

Specifying the change in the coordinate system with Euler angles by rotating
around the z-axis by $\phi\in[0,2\pi]$, then around the new y-axis by angle
$\theta\in[0,\pi]$ and finally around the new z-axis by angle
$\gamma\in[0,2\pi]$, the harmonic coefficients transform as (see for example
\cite{VarMosKhe88})
\begin{equation}
\tilde t_{\ell \tilde m} = \sum_{m}  t_{\ell m} D_{m \tilde m
}^\ell(-\gamma,-\theta,-\phi)\,,
\slabel{eqn:trot}
\end{equation}
where the Wigner rotation matrix is given by
\beq
D_{m \tilde m }^\ell(\gamma,\theta,\phi)= {\rm e}^{-i \tilde m \gamma} d_{m
  \tilde m}(\theta) {\rm e}^{-i  m \phi}
\eeq
and
\begin{eqnarray}
&& d_{m  {\tilde m}}(\theta) =   \nonumber \\
&&\sum_{k={\rm max}\{0,m+{\tilde m}\}}^{{\rm min}\{\l+m,\l+{\tilde m} \}}
	 \left(\cos{\theta\over 2}\right)^{2k-m-{\tilde m}} 
	 \left(\sin{\theta\over 2}\right)^{2\l+m+{\tilde m}-2k}\nonumber\\
&&	\times (-1)^{k+\l+m} {\sqrt{(\l+m)!(\l-m)!(\l+{\tilde m}
	     )!(\l-{\tilde m} )!} 
	\over k!(\l+m-k)!(\l+{\tilde m} -k)!(k-m-{\tilde m} )!} \nonumber
\end{eqnarray}
is a real function.
   Thus the ensemble average becomes
\begin{equation}
\langle \tilde  t_{\ell \tilde m}^* \tilde t_{\ell' \tilde m'}\rangle
= \sum_m \langle  t_{\ell m}^* t_{\ell' m }\rangle D_{ m \tilde
  m}^{\ell*} D_{ m \tilde m'}^{\ell} \,,
\slabel{eqn:covrot}
\end{equation}
and exhibits completely broken isotropy.

\begin{table}
\caption{\footnotesize \label{tab:tlms} 
WMAP data vs. fiducial model in the dipole vs maximum angular
momentum direction (see \S \ref{sec:proofofprinciple}).  $N_\ell/C_\ell^{\rm fid}$ shows
the relative contribution of noise assumed.}
\begin{tabular}{llllll}
\hline \hline
 $\ell$& $m$ & $C_{\ell}^{\rm fid} (\mu$K$^{2})$ & $N_\ell/C_\ell^{\rm fid}$ &
\multicolumn{2}{c}{$|t_{\ell m}|^2/C_\ell^{\rm fid}$} \\
&&&
&  dipole &  $\hat L^2_{\rm max}$ \\
    2 &    0 & 1233 &      0.005 &      0.005 &      0.005 \\ 
     &    1 &  &      &      0.013 &      0.003 \\ 
     &    2 &  &      &      0.396 &      0.406 \\ 
    3 &    0 &  577 &      0.007 &      0.179 &      0.315 \\ 
     &    1 &  &       &      0.079 &      0.003 \\ 
     &    2 &  &       &      0.426 &      0.029 \\ 
     &    3 &  &       &      2.155 &      2.560 \\ 
    4 &    0 &  322 &      0.009 &      3.929 &      1.641 \\ 
     &    1 &  &       &      0.156 &      0.939 \\ 
     &    2 &  &       &      0.052 &      0.508 \\ 
     &    3 &  &       &      0.271 &      0.401 \\ 
     &    4 &  &       &      0.406 &      0.180 \\ 
    5 &    0 &  202 &      0.011 &      0.035 &      0.014 \\ 
     &    1 &  &       &      0.376 &      0.811 \\ 
     &    2 &  &       &      0.035 &      1.254 \\ 
     &    3 &  &       &      6.827 &      2.777 \\ 
     &    4 &  &       &      0.077 &      2.769 \\ 
     &    5 &  &       &      0.364 &      0.078 \\ 
\hline
\hline
\end{tabular}
\end{table}

In Tab.~\ref{tab:tlms} and Fig.~\ref{fig:wmap}, we compare the power in the
observed multipole moments of WMAP in the dipole frame to $C_\ell^{\rm fid}$,
the power spectrum of a fiducial isotropic $\Lambda$CDM model.  This model,
which we use throughout the paper, has
$\Omega_{m}=1-\Omega_{Q}=0.27$, 
$h=0.72$, $\Omega_{b}h^{2}=0.024$, $\delta_{\zeta}= 5.07
\times 10^{-5}$ and $\tau=0.17$.  The WMAP data is
taken to be the cleaned full-sky map of Tegmark et al.\ (hereafter TOH;
\cite{TOH}).  Note that the $m=0$ components of the quadrupole and octopole are
low both with respect to the average over $m$ in the WMAP data and compared
with the fiducial model.  Thus the power {\it deficit} in the quadrupole and
octopole are aligned.  With broken statistical isotropy it is in principle
possible to explain these and other anomalies in the power distribution of the
data. 

By employing the TOH cleaned map we are ignoring residual foregrounds and
systematics.  While these may or may not weaken the significance of alignments
in the data \citep{Bennett_foregr, LILC, SS2004,Bielewicz2004, Copi2005}, they
do not affect the ability of a cosmological model to produce them.  The latter
is the main topic we address in this work.

\section{Additive Isotropy Breaking}
\label{sec:additive}

To illustrate the idea of spontaneous isotropy breaking we first take the
simple case of an additive contribution to the anisotropy.  Spontaneous
isotropy breaking requires a non-linear response in the observed CMB
temperature to an underlying gradient field.  We begin with a motivating
example of a known systematic effect, the non-linear response of a detector, in
\S \ref{sec:detector}.  This non-linear response modulates the CMB dipole into
higher order anisotropy aligned with the dipole.  We then apply this idea to a
cosmological example involving dark energy density fluctuations in \S
\ref{sec:darkenergy}.

The statistical anisotropy caused by additive isotropy breaking generally produces
alignments of {\it excess} power.  While alignments of power deficits,
like those seen in the quadrupole and octopole of the WMAP data, are possible
we show in \S \ref{sec:excess} that they are statistically {\it less likely}
than a chance alignment in a statistically isotropic field.  This conclusion
holds true for any mechanism which seeks to restore power and hence statistical
isotropy in the quadrupole and octopole by removing a contaminant that is
uncorrelated with the intrinsic anisotropy.

\subsection{Instrumental Example}
\label{sec:detector}

A concrete but non-cosmological means of additive isotropy breaking is provided
by the non-linearity of a detector responding to temperature fluctuations.
Suppose that the detector responds to a true temperature 
signal of $G(\hat{\bf n})$ as
\begin{eqnarray}
T_{\rm det}(\hat{\bf n}) &=& f F[G(\hat{\bf n})] =f  \sum_i \alpha_i 
\left[ {G(\hat{\bf n}) \over f }\right] ^i  \,,
\end{eqnarray}
 where $\alpha_1=1$. 
Here $f$ is an arbitrary normalization scale for the
non-linearity of the response.  If $\alpha_{i>1} \ne 0$ then $T_{\rm det} \ne G$
and the observed temperature is a non-linear modulation of the true temperature.

The dominant temperature signal in a differencing experiment is the dipole
arising from our peculiar motion, $G(\hat{\bf n})=T_{\rm dip}\cos\theta$ in
the dipole frame, with $T_{\rm dip}=3.35$mK.  The non-linear response has
the general form of Eqn.~(\ref{eqn:generalform}). Taking $f=T_{\rm dip}$,
\begin{eqnarray}
{T_{\rm det}(\hat{\bf n})\over T_{\rm dip}} &=& 
\alpha_1 P_1(\cos\theta) + \alpha_2 \left [{2 \over 3}P_2(\cos\theta) +1\right ] 
   \\ 
&&  + \alpha_3 \left [{2\over 5} P_3(\cos\theta) + {3 \over 5} P_1(\cos\theta)\right ] 
+ \ldots \,, \nonumber
\end{eqnarray}
where $P_\ell =\sqrt{4\pi/(2\ell+1)} Y_{\ell 0}$ are the Legendre polynomials, and therefore
\begin{eqnarray}
 w_2 &=& {4\over 3} \sqrt{\pi \over 5} \alpha_2 + \ldots  \,,  \nonumber \\
 w_3 &=& {4\over 5} \sqrt{\pi \over 7} \alpha_3 + \ldots  \,,
\end{eqnarray} 
where $\ldots$ represent contributions from $i>3$.   With $\alpha_2 = \alpha_3 \sim 10^{-2}$,
the $10^{-3}$ dipole anisotropy is modulated into a $10^{-5}$ quadrupole and octopole
anisotropy with $m=0$ in the dipole frame.

Given the observed multipole moments $t_{\ell m}$, the intrinsic anisotropy becomes
\begin{eqnarray}
a_{\ell m} = t_{\ell m} - f w_{\ell} \delta_{m0} \,.
\end{eqnarray}
For multipole moments that are observed to be anomalously low, as is the case
for the quadrupole and octopole moments of WMAP in the dipole direction (see
Tab.~\ref{tab:tlms}), $t_{\ell m} \rightarrow 0$ and $|a_{\ell 0}|^2
\rightarrow f^2 w_{\ell}^2 \delta_{m 0}$.  Thus the removal of the additive
contribution {\it restores} power to the intrinsic sky and can bring it closer
to a statistically isotropic distribution.  Note that this restoration involves
a chance cancellation between a Gaussian random realization of $a_{\ell m}$ and
a fixed additive contribution $f w_{\ell}$.  This will be important in
addressing the probability of such an occurrence in \S \ref{sec:excess}.

Detector non-linearity is unlikely to be the actual source of the observed
alignments.  In this model, the nonlinearity of the response would be at the
percent level for the $3.35$mK dipole.  If this model were extrapolated to
brighter sources, it would imply that $\sim 1$K sources completely saturate the
response which is demonstrably not true for the WMAP instrument.  Nonetheless,
the non-linear detector response model provides a concrete and conceptually
simple illustration of the isotropy breaking mechanism. It has the added benefit
that it naturally picks out the dipole direction as the preferred axis along which 
the temperature is modulated as seen in the WMAP data.

\subsection{Dark Energy Example}
\label{sec:darkenergy}

Cosmological examples of isotropy breaking follow a similar form but are more
complicated to calculate.  Here one must combine the dynamics of how the
modulated field produces temperature fluctuations in physical space along with
the projection along the line of sight onto angular space.  The benefit of a
cosmological model is that the spectrum $w_\ell$ of the additive contributions
have a well defined shape and hence the model in principle has predictive
power, e.g. an additive correction at relatively high multipoles predicts a
specific change to the low order multipoles and vice versa.

As a concrete example, let us consider the Integrated Sachs-Wolfe (ISW) effect
induced by spatial perturbations in the dark energy density from a quintessence
field $Q$.  If the perturbations to the field $Q$ are dominated by superhorizon
scale fluctuations then within our horizon the field will look linear
\begin{equation}
Q_i = B_0 + B_1 x_3 + \ldots \,,
\end{equation}
where $B_i$ are constants, so that the local gradient picks out a preferred axis
${\bf x}_3$.  The non-linear modulation is provided by the scalar field potential.
Taking it to be of the form
\begin{equation}
V(Q) = V_0 [1+ f \cos(Q/M_0)] \,,
\end{equation}
the field gradient manifests itself as
\begin{equation}
V(Q_i) = V_0[1 + f \cos(k_0 x_3 + \delta)] \,,
\end{equation}
where $k_0 \equiv B_1/M_0$ and $\delta \equiv B_0/M_0$.
This initial potential energy fluctuation is then transferred onto the CMB through
the ISW effect and the dynamics of the field.

As the actual field gradients are from superhorizon wavelengths, the field will remain
frozen provided its mass $m$ is small compared to the Hubble parameter $H$:
\begin{eqnarray*}
{|m^2|\over H^2}=
\left| { \partial^2 V \over \partial Q^2} \right| 
{1 \over H^2} & \le & 3 \left| {\partial^2 V \over
\partial Q^2}\right| {\Mp^2\over V} \le 3 f\left( {\Mp \over M_0 }\right)^2\,,
\end{eqnarray*}
where $M_{\rm pl} = (8\pi G)^{-1/2}$ is the reduced Planck mass.
Thus, if 
\begin{equation}
{M_0 \over M_{\rm pl}} \gg f^{1/2} \,,
\end{equation}
the field will not roll and the density field then carries potential energy
fluctuations that are frozen.  Its Fourier components are
\begin{eqnarray}
{\delta \rho_Q \over \rho_Q}({\bf k}) &= &{f \over 2}e^{i\delta}
 (2\pi)^3 \delta({\bf k} - k_0 \hat{\bf x}_3) \nonumber\\
 &&+
{f \over 2}e^{-i\delta} (2\pi)^3 \delta({\bf k} + k_0 \hat{\bf x}_3)
 \,. \slabel{eqn:deltarhoQ}
\end{eqnarray}

\begin{figure}
\epsfig{file=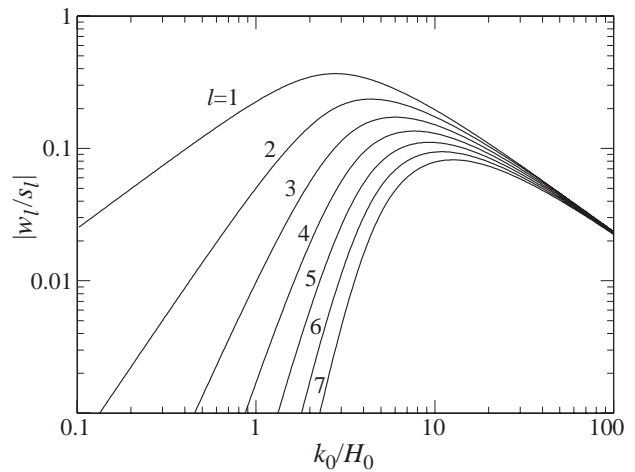, width=3.25in}
\caption{\footnotesize Dark energy based additive contributions $w_\ell$.
Shown are the first few multipoles of $|w_{\ell }/s_{\ell}|$ as a function of
$k_0/H_0$.  Long wavelength perturbations exhibit a spectrum in $w_\ell$ that
is steeply falling in $\ell$ and is modulated by parity considerations from
$s_{\ell}$.}
\label{fig:isw_alm}
\end{figure}

For the moment, we consider the case where only the dark energy is initially
perturbed. The comoving curvature perturbation, $\zeta$, will evolve if there
is a perturbation to the total pressure, $p_T$. As only the dark energy is
perturbed, initially the Universe is effectively homogeneous, $\zeta_i=0$, and
$\zeta$ only begins to grow during dark energy domination when the only
non-negligible components to the total energy density $\rho_T$ are the dark
energy density and the pressureless matter, $\rho_m$. The dark energy is
assumed to be sufficiently light that it is approximately frozen and so $\rho_Q
\approx -p_Q =$ const. The evolution of the comoving curvature perturbation, in a flat
Universe, is given by (see for example the Appendix of \cite{GorHu04})
\begin{eqnarray}
\zeta &=& \zeta_i - \int_0^a {da' \over a'} {\delta p_T \over {\rho_T
    +p_T}} \nonumber \\
&\approx& \int_0^a {da' \over a'}{\delta \rho_Q \over \rho_m}\, . \slabel{eqn:zeta}
\end{eqnarray}
Denoting the relative energy densities of dark energy and matter today 
as $\Omega_Q\equiv \rho_Q / \rho_T |_{a=1}$ and $\Omega_m\equiv \rho_m / \rho_T
|_{a=1}$, the
usual redshifting of pressureless matter can be written as
\begin{equation*}
\rho_m = {\rho_Q \over a^3} {\Omega_m \over \Omega_Q}\, .
\end{equation*}
Substituting this into \eq{eqn:zeta} gives
\begin{equation}
\zeta = {a^3 \over 3} {\delta \rho_Q \over \rho_Q} {\Omega_Q \over
  \Omega_m}\, .
\slabel{eqn:zeta1}
\end{equation}
The Newtonian gravitational potential can be expressed in terms of
$\zeta$ as (see for example \cite{HuEis98,GorHu04})
\begin{eqnarray}
-\Psi({\bf k},a) &=&  \zeta - {H \over a} \int_0^a {da' \over H} \left (\zeta
-{\delta p_T \over \rho_T + p_T}\right ) \nonumber \\
&\approx& \zeta - {H \over a} \int_0^a {da' \over H} \left (\zeta
+{\delta \rho_Q \over \rho_m}\right )\, .
\end{eqnarray}
Substituting  \eqs{eqn:zeta}{eqn:deltarhoQ} into this gives
\begin{equation}
\Psi({\bf k},a) \approx \psi  (2\pi)^3\delta({\bf k} - k_0 \hat{\bf x}_3)  + \psi^*(2\pi)^3
 \delta({\bf k} + k_0 \hat{\bf x}_3) \,,
\end{equation}
where
\begin{equation}
\psi  = -{1 \over 3}{\Omega_Q \over \Omega_m}   {f \over 2}e^{i\delta} \left( a^{3}-4
{H(a) \over a} \int {da' \over H(a') } {a'}^3\right)\, .  
\end{equation}

When there are only dark energy perturbations, the temperature fluctuations in
the direction $\hat{\bf n}$ due to the potential $\Psi$ are given by the ISW
effect.  Relating these contributions to the form of an additive isotropy
breaking term in \eq{eqn:generalform} gives
\begin{eqnarray}
{\Delta T(\hat{\bf n}) \over T} = f w(\hat{\bf n}) = \int 2 \Psi'({\bf x}=D
\hat{\bf n},a)d\ln a\,,
\end{eqnarray}
where $D$ is the comoving distance along the line of sight and primes are
 derivatives with respect to $\ln a$.
The multipole moments are given by
\begin{equation}
 f w_{\ell} \equiv  \int d\hat {\bf n} Y_{\ell m}^{*}({\hat{\bf n}})
  {\Delta T \over T}(\hat{\bf n}) \, .
\end{equation} 
With the Rayleigh expansion of a plane wave 
\begin{equation}
\exp(i {\bf k} \cdot {\bf x}) = \sum_{\ell m} 4\pi i^{\ell} j_{\ell}(k
D) Y_{\ell m}^{*}({\hat{\bf k}}) 
Y_{\ell m}({\hat{\bf n}}) \,,
\end{equation}
they can be written as 
\begin{equation}
 f w_{\ell}= \int d\ln a \int{d^{3}k \over (2\pi)^{3}} 4\pi i^{\ell} j_{\ell}(k
D) Y_{\ell m}^{*}({\hat{\bf k}}) 
 2\Psi' ({\bf k},a)\,.
\end{equation}
Taking the form of the Newtonian gravitational potential above, we get
\begin{eqnarray}
w_\ell &= &
-{\Omega_{Q} \over {\Omega_m}} s_\ell \sqrt{4\pi(2\ell+1)} \int 
d\ln a j_\ell(  k_0 D ) I(a) a^3\,, \nonumber\\[0.1cm]
s_\ell & \equiv &
  \cos\delta  (-1)^{\ell/2}\delta_{\ell}^{\rm e} +
\sin\delta (-1)^{(\ell+1)/2}  \delta_{\ell}^{\rm o}\,,
\end{eqnarray}
where $\delta_{\ell}^{\rm e} = 1$ if $\ell$ is even and 0 if $\ell$ is odd, and
vice versa for $\delta_{\ell}^{\rm o}$.  Here we have defined
\begin{eqnarray}
I(a) \equiv -{8 \over 3 a^3} {d \over d\ln a} \left[ {H(a) \over a } \right]
\int {d a' \over H(a')} {a'}^3 - {2 \over 3}\,.
\end{eqnarray}
Given that the CMB response to a dark energy {\it density} perturbation is still
linear, restoring adiabatic fluctuations is a simple matter of adding in the
fiducial model contributions as an uncorrelated contribution to the temperature field.

\begin{figure}
\epsfig{file=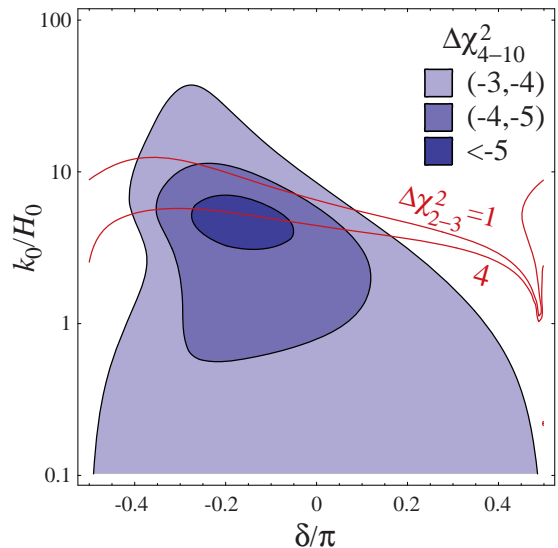, width=3in}
\caption{\footnotesize $\Delta\chi^{2}$ relative to the fiducial model
for the additive model parameter space of phase 
and wavelength $(\delta,k_{0}/H_{0})$ with $f$ chosen to minimize the
$\Delta\chi^{2}$.  For the multipoles $\ell=4-10$, the maximum improvement is
$\Delta\chi^{2}_{4-10}=-5.4$ (shaded contours).  These models restore
power to the quadrupole and octopole and thus $\ell=2, 3$ add a disfavored 
positive
contribution through $\Delta\chi^{2}_{2-3}$ (curves).}
\label{fig:model_par_space}
\end{figure}

In Fig. \ref{fig:isw_alm}, we show the amplitude of the first few multipoles of
$w_\ell/s_\ell$ as a function of $k_0/H_0$ in the fiducial cosmological model.
There are two general features in $w_\ell$ that go beyond the specifics of this
model. First, a spatial modulation projects onto an angular modulation with a
weight given by the spherical Bessel function $j_\ell$.  For a superhorizon
fluctuation
\begin{equation}
k_0 D = {k_0 \over H_0} H_0 D \sim {k_0 \over H_0} \ll 1 \,,
\end{equation}
and so $j_\ell \propto (k_0/H_0)^\ell$.  Thus a superhorizon scale modulation
typically predicts a sharply falling spectrum in $\ell$ for the additive
contribution.  The modulation scale $k_0$ should not be confused with the
field gradient scale which is by assumption always superhorizon. 

The symmetry factor $s_\ell$ provides an interesting
modulation of this result, and this is the second general feature of the model.
The phase of the plane wave $\delta$ determines its symmetry under a parity
transformation $\hat {\bf n} \rightarrow -\hat{\bf n}$ or $\hat {\bf x_3}
\rightarrow -\hat{\bf x_3}$.  Because spherical harmonics have definite parity
$(-1)^\ell$, the plane wave modulation affects even and odd multipole moments
differently.  In particular choosing a phase $\delta \rightarrow 0$, eliminates
the contributions from all odd multipoles.

\begin{figure}
\epsfig{file=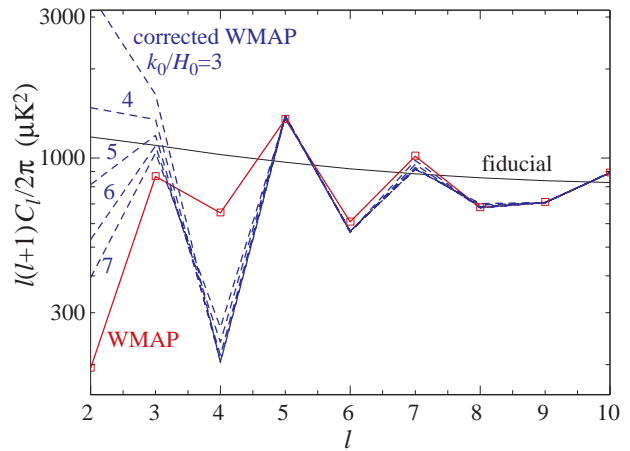, width=3.25in}
\caption{\footnotesize Restoration of power in the dark energy additive model.  Taking a phase
of $\delta=-0.173\pi$ that intersects the minimum $\Delta\chi_{4-10}^{2}$ 
from Fig.~\ref{fig:model_par_space}, models with wavelengths $k_{0}/H_{0}$ around
the minimum restore power  to the quadrupole and octopole (dashed line
vs points) and bring them closer to the fiducial model (curve).}
\label{fig:corrected}
\end{figure}

\subsection{Power Excess and Restoration}
\label{sec:excess}

Given the dark energy based model of the previous section, we can test the WMAP
data for additive contributions that break statistical isotropy.  We set the
preferred axis for isotropy breaking to be in the direction of the dipole based
on the observed pattern of low multipole alignments
(e.g. \cite{deOliveira2004,Schwarz2004,Land2005a}).  In the dipole frame, the observed
power in $t_{\ell m}$ is given as a fraction of the power in the fiducial model
of Fig.~\ref{fig:wmap} in Tab.~\ref{tab:tlms}.  The uneven distribution of
power across $m$ for the quadrupole and octopole quantifies the alignment
problem.

To determine whether the WMAP data favor an additive breaking of statistical
isotropy, we maximize the likelihood with respect to model parameters.  Since
the additive contribution is deterministic and the underlying temperature field
is assumed to be drawn from the fiducial model this amounts to minimizing the
$\chi^2$
 \begin{equation}
 \chi^2 = \sum_{\ell_{\rm min}}^{\ell_{\rm max}} (C_\ell+N_\ell)^{-1}
 ( t_{\ell m}^* - f w_{\ell}\delta_{m0})  ( t_{\ell m} - f w_{\ell}\delta_{m0}) \,.
 \end{equation}
 Here we have included a crude estimate of the noise in the WMAP data from
 their quoted power spectrum errors in the absence of sample variance.  These
 are listed in Tab.~\ref{tab:tlms}.  Note that our results here are insensitive
 to the specific representation of the noise as it is much smaller than the rms
 signal of the fiducial model.
 
 The benefit of having a cosmological model for the additive contributions is
 that the functional form of the contributions is determined through $w_\ell$.
 Evidence for an alignment of excess power in one range of multipoles predicts
 the correction that must be applied to another.  We therefore minimize the
 $\chi^2$ for $4 \le \ell \le 10$ with respect to the parameters $(f,
 k_0/H_0,\delta)$ and then examine what it implies for the quadrupole and
 octopole.  The best-fit solution is $k_0/H_0=4.8, \delta=-0.173\pi, f=1.16\cdot
 10^{-4}$ and has improvement of $\Delta \chi^2=-5.4$ relative to the fiducial
 model with no additive contribution.  In fact, there is a wide range of models
 with a similar level of improvement ($\Delta \chi^2\sim -5$) since the two
 parameters $k_0/H_0$ and $\delta$ affect the $\ell$ dependence of the
 contributions similarly.  We show in Fig.~\ref{fig:model_par_space} the
 $\Delta \chi^2$ contours for $\ell = 4-10$ in the ($k_0/H_0$, $\delta$) space
 for the best-fit amplitude $f$.  About two thirds of the improvement in
 $\chi^2$ comes from $\ell = 4$ which has a $\sim 2\sigma$ excess in power
 compared with the fiducial model (see Fig.~\ref{fig:wmap} and
 Tab.~\ref{tab:tlms}).

Given this range of models that improve the $\Delta\chi^2$ for $\ell=4-10$, it
is interesting to see what they imply for the quadrupole and octopole.  In
Fig.~\ref{fig:corrected}, we show the intrinsic WMAP power spectrum (i.e.\ one
with the additive contribution removed)
\begin{equation}
{1\over 2\ell  +1}\sum_m | t_{\ell m} - f w_{\ell} \delta_{m0}|^2
\end{equation}
for $\delta =-0.173\pi$ and a range of $k_0/H_0$
that includes the best fit model.  Models around the best fit restore power
to the quadrupole and octopole to bring them closer to the fiducial model.

There are however a number of problems that make this explanation of the
alignment problem unsatisfactory.  First, although the removal of the excess
power in $\ell=4$, $m=0$ brings the distribution in $m$ of the power closer to
flat, the total power is well under the fiducial model.  Our additive model is
limited in that it can only affect $m=0$ in the preferred frame.  Similarly the
quadrupole and octopole multipole moments are low not only in $m=0$ but also
$m=\pm 1$ (see Tab.~\ref{tab:tlms}) which cannot be altered in our additive
model.

The most important flaw however is that to restore power to the quadrupole and
octopole one must assume that the additive contribution nearly exactly cancels
the intrinsic temperature fluctuation.  To assess whether that is a likely
occurrence, consider the $\Delta\chi^2$ contributed by $\ell=2$ and $3$.
Around the minimum for $\ell =4-10$, it receives a positive contribution of
$\Delta\chi^2_{2-3}$ between 1 and 4 relative to the fiducial model (see
Fig.~\ref{fig:model_par_space}).  A chance cancellation of power is {\it less}
likely than a realization of anomalously low power.  For example,
in random realizations of the fiducial model without the additive correction we
find 2.3\% of the realizations have a lower average power in the quadrupole
moments than WMAP.  With the best fit additive correction this fraction drops
to 0.8\%.

Note that quadrupole and octopole problem remains even had the improvement from
$\Delta \chi^2_{4-10}$ been significantly larger.  It  also remains if the figure of
merit were the alignment of the quadrupole and octopole and not $\chi^2$. The
reason is that the Gaussian distribution for the intrinsic sky in $a_{\ell m}$
peaks at zero.  The probability that a realization of $a_{\ell 0}$ falls within
$\delta a_{\ell 0}\sim |t_{\ell 0}|$ of the additive contribution $w_\ell$ is
uniformly lower than the probability that it fell by chance at a $|a_{\ell 0}|<
\delta a_{\ell 0}$ without $w_\ell$.  The same argument holds for alignments of
power deficits since they are formed from the joint probability of low power
across a range of $\ell$.

These conclusions hold for any additive model where a template contamination by
chance cancels the intrinsic fluctuations. This includes explanations of the
alignments involving foregrounds and broken isotropy in the background metric
as in Bianchi models \cite{Jaffe2005}.  A cancellation model can at best be
said not to decrease the probability of alignments of low power multipoles by a
substantial amount.  It can never increase the probability.

There is one remaining possibility for additive models.  An additive
contribution that is not azimuthally symmetric around the preferred axis might
alone account for essentially all of the observed quadrupole and octopole.  For
example, foreground contamination could in principle supply the observed power.
%
Unfortunately, such an explanation of alignments
would then exacerbate the large angle power deficit problem.

\begin{figure}
\epsfig{file=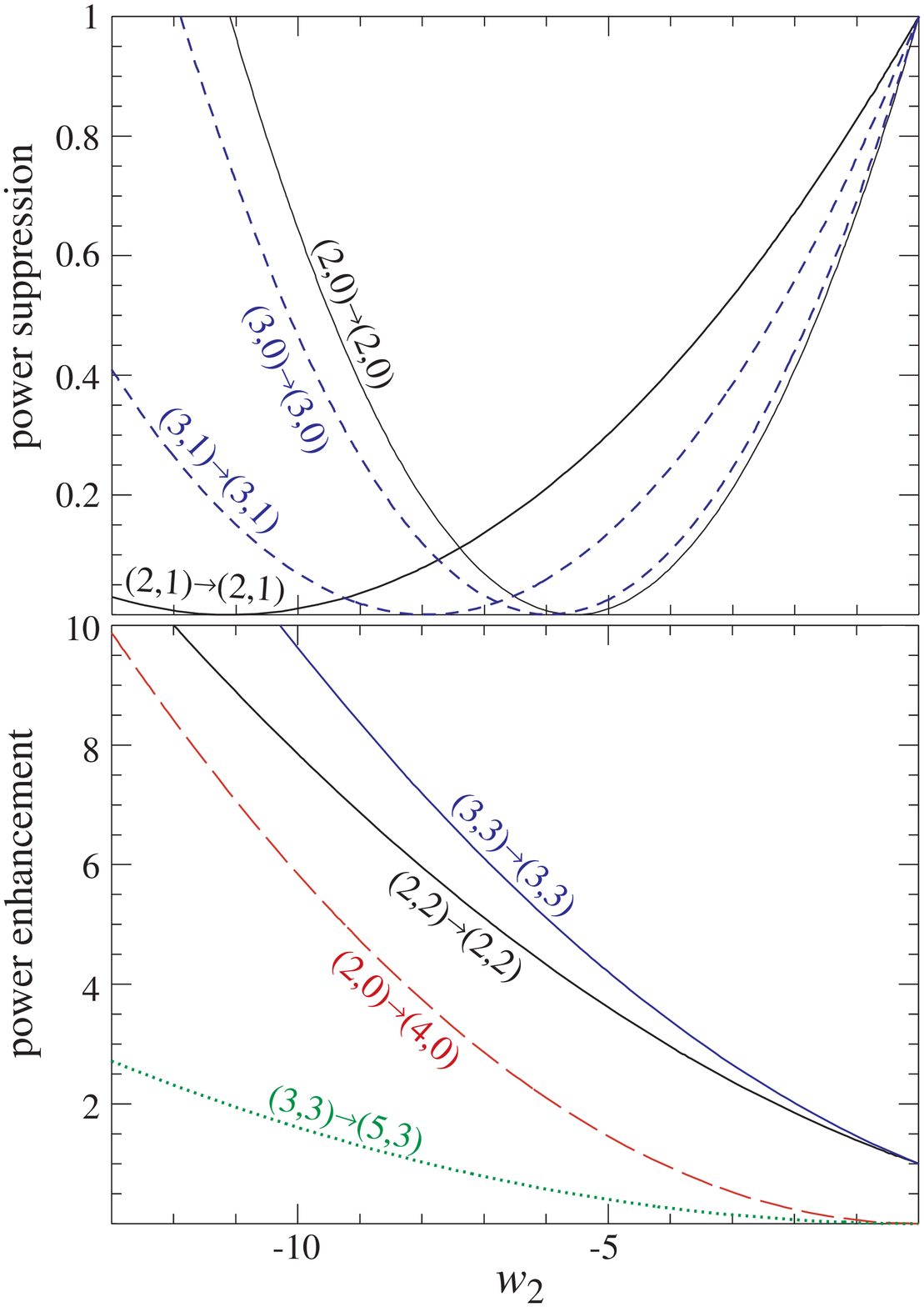, width=3.25in}
\caption{\slabel{fig:magicw2}  
\footnotesize
Multiplicative model with a quadrupolar modulation.  Shown is
the fractional change or transfer of power from an intrinsic multipole $(\ell,m)$ to
an observed $(\ell',m')$ or $p_{(\ell,m) \rightarrow (\ell',m')}^2$.
For a modulation amplitude $-5 \simlt w_{2} \simlt -10$, power in $\ell=2,3$
and $m=0,1$ is suppressed while that in $(\ell,m) = (2,2)$, $(3,3)$, $(4,0)$ and
$(5,3)$ is enhanced.   These features are seen in the WMAP TOH data of 
Tab.~\ref{tab:tlms} and enhance the planarity and alignment of the quadrupole
and octopole.}
\label{fig:magic}
\end{figure}

\section{Multiplicative Isotropy Breaking}
\label{sec:multiplicative}

A multiplicative model for breaking statistical isotropy eliminates the two
fundamental flaws of the additive model.  Firstly, though the modulation is
still azimuthally symmetric it affects multipole moments of all $m$ values.
Secondly, the change in the observed multipole moments is correlated with the
intrinsic anisotropy and hence can explain alignments of power deficits.
Together these features can explain the planarity and alignment of the
quadrupole and octopole seen in WMAP.

Based on the instrumental example of a calibration error that varies across the
sky (\S \ref{sec:calibration}), we construct a proof of principle model (\S
\ref{sec:proofofprinciple}) that solves the quadrupole and octopole  mutual alignment
problem (\S \ref{sec:alignments}).

\subsection{Instrumental Example}
\label{sec:calibration}

An instrument whose calibration varies across the sky will modulate the
intrinsic anisotropy $B(\hat{\bf n})$ as
\begin{equation}
T(\hat{\bf n}) =  f  [1+ w(\hat{\bf n}) ] B(\hat{\bf n}) \,,
\end{equation}
where $f$ is the sky averaged calibration and $w$ is the fluctuation in the
calibration.  A calibration fluctuation can transfer power from one multipole
moment to another as well as alter the power in a given multipole.

Consider for a calibration fluctuation that is purely
quadrupolar in nature 
\begin{equation}
w(\hat{\bf n})=w_2 Y_{20}(\hat{\bf n})\,.
\label{eqn:quadmod}
\end{equation}
This form is motivated by the need to suppress fluctuations along both poles
of the preferred axis.  It also relates multipoles of $\Delta \ell=2$ by 
\eq{eqn:covmatrix}  (see also \cite{Pruetal04} for a dipolar example) 
which as we will see has interesting implications
for alignments and parity anomalies beyond the quadrupole and
octopole \cite{Land2005a,Land2005c}.

The effect of this calibration modulation is an $\ell$ and $m$ dependent
alteration of the amplitude of the temperature field as well as a transfer of
fluctuations between $\Delta \ell=2$ modes of the same $m$.  From \eq{eqn:tlm},
the contribution from an intrinsic fluctuation $b_{\ell'm'}$ to an observed
multipole moment $t_{\ell m}$ is given by
\begin{equation}
 f b_{\ell' m'}p_{(\ell',m')\rightarrow(\ell,m)} \,,
\end{equation}
where
\begin{eqnarray}
p_{(\ell',m')\rightarrow(\ell,m)}
&=& \left( \delta_{\ell \ell'}+ R_{\ell m}^{2\, \ell'}\right) \delta_{mm'}\,.
\end{eqnarray}
These square of several of these $p$'s are plotted as a function of $w_2$ in
Fig.~\ref{fig:magic}.  For certain values of $w_2$, the modulation can suppress
specific multipoles of the quadrupole and octopole to zero.  Interestingly for
$-10 \simlt w_2 \simlt -5$, the suppression is efficient for $m=0,\pm 1$.
Moreover for this range, the power in $\ell=2$, $m=\pm 2$ and $\ell=3$, $m=\pm
3$ is actually enhanced.  These properties are exactly what is needed to
explain the observed power distribution of the WMAP sky.

Furthermore, a modulation in this regime also enhances power in $\ell =4$,
$m=0$ from intrinsic power in $\ell=2$, $m=0$ and $\ell=5$, $m=3$ from
$\ell=3$, $m=3$.  The WMAP multipoles $(4, 0)$ and $(5, 3)$ 
have excess
power in the dipole direction.  Finally the modulation
correlates the observed signal between these pairs of multipoles.  The
correlation for $\ell=2$, $m=0$ is not very relevant since its observed value
is consistent with noise (see Tab.~\ref{tab:tlms}).  For $\ell=3$, $m=3$ it is
of the right sign to account for enhanced fluctuations in both multipoles as
seen in the WMAP data \cite{Land2005a}.

Unfortunately a simple calibration error cannot literally be the cause of
alignments.  In the discussion above we have only considered the modulation of
intrinsic power in the quadrupole and octopole.  A calibration error would
modulate all of the intrinsic anisotropy.  In this case terms such as $p_{(4,0)
\rightarrow (2,0)}$ would generate observed quadrupole power out of intrinsic
$\ell=4$ fluctuations and spoil the ability to lower its power.

\begin{figure}
\epsfig{file=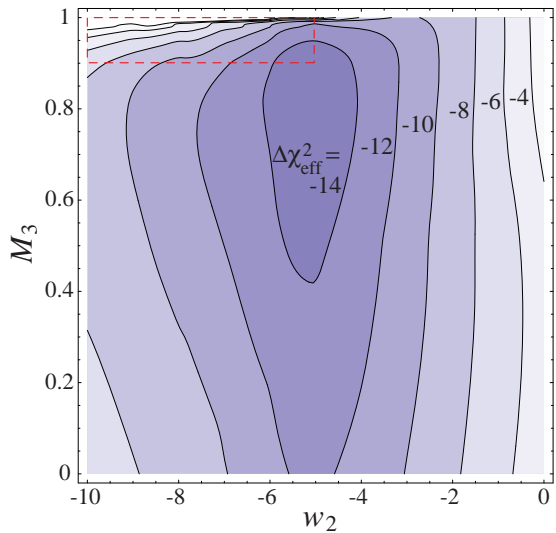, width=3in}
\caption{ \footnotesize \slabel{fig:chi2eff_dip} $\Delta\chi_{\rm eff}^2 \equiv
-2 \Delta \ln {\cal L}$ for the multiplicative model with the dipole as the
preferred axis. Models with $w_{2}\sim -5$ and $M_{3} \approx 0.8$ have
$\Delta\chi_{\rm eff}^{2}\approx -16$ and best match the dipole frame data
given that the most anomalous power deficit is in $\ell=2$, $m=0$ (see
Fig.~\ref{fig:magicw2}).  The amplitude $f$ is set to minimize $\Delta
\chi_{\rm eff}^{2}$ for each model.  A range of models around the maximum
improve the $\Delta\chi_{\rm eff}^{2}$ and most of the improvement is due to
the breaking of statistical isotropy through the single parameter $w_{2}$. The
dashed box highlights models which best produce angular momentum alignments in
Fig.~\ref{fig:alignments_dip}.}
\end{figure} 

\begin{figure}
\epsfig{file=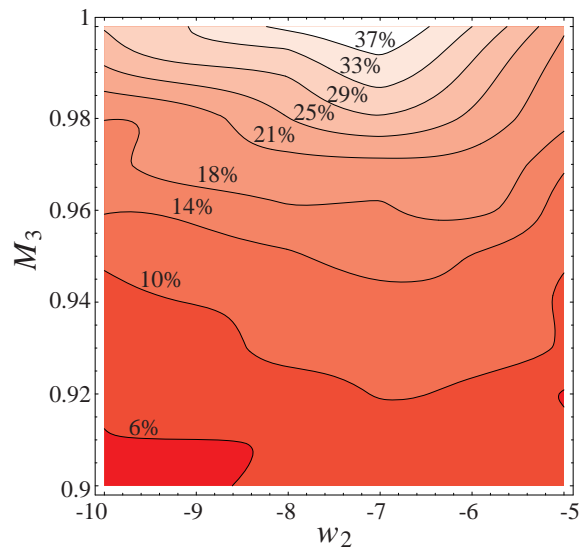, width=3.15in}
\caption{\footnotesize
\slabel{fig:alignments_dip} Percentage of multiplicative models that
exhibit a higher aligned angular momentum statistic $ \hat L^2$ for the quadrupole
and octopole than WMAP.  Here
$f$ is chosen to minimize the $\Delta\chi^{2}_{\rm eff}$ of the data in the dipole
frame.  The percentage peaks at 45\% for $w_{2}\approx -7$ and $M_{3}=1$ and is
to be compared with 0.22\% for the fiducial model.}
\end{figure} 

\subsection{Proof of Principle Example}
\label{sec:proofofprinciple}

At least in principle, a cosmological version of the calibration mechanism has
the ability to modulate only the quadrupole and octopole.  Large scale
temperature anisotropies can arise from both the Sachs-Wolfe and integrated
Sachs-Wolfe effect and each can carry extra degrees of freedom from
isocurvature fluctuations or multiple sources of curvature fluctuations.  It is
possible then that the presence of a long wavelength field gradient modulates
only one of these effects.

Constructing a fully physical model is beyond the scope of this paper.  Here we
seek only a proof of principle that a multiplicative modulation of the
intrinsic anisotropy can explain the WMAP anomalies. This proof of principle
should help to guide the construction of an physical model that satisfies the
observational requirements.

Let us take the general form for the modulation \eq{eqn:generalform} where the
anisotropy is constructed out of two contributions $A(\hat{\bf{n}})$ and
$B(\hat{\bf{n}})$, only one of which is modulated.  Based on the discussion of
calibration error in the previous section, we parameterize their underlying
power spectra as
 \begin{eqnarray}
C_\ell^{aa} &=& 
(1-M_\ell)C_\ell^{\rm fid}\,, \nonumber\\
C_\ell^{ab} &=& 0\,, \nonumber\\
C_\ell^{bb} &=& 
                                         M_\ell C_\ell^{\rm fid} \,.      
                                         \label{eqn:toymodel}         
 \end{eqnarray}
Here $C_\ell^{\rm fid}$ is the fiducial cosmological model specified in
\S\ref{sec:darkenergy}.  We will further take the modulation to be purely
quadrupolar in nature as given by \eq{eqn:quadmod}.  As discussed in \S
\ref{sec:darkenergy}, a nearly pure quadrupolar modulation can be achieved in
physical space by the combination of a long wavelength spatial modulation whose
phase corresponds to a nearly even parity when projected on the sky.
 
The multiplicative modulation model is then specified by three quantities:
$w_2$ which controls the relative amplitude of the modulation in the field
$B(\hat{\bf{n}})$, $f$ which describes the relative amplitude of
$A(\hat{\bf{n}})$ and $B(\hat{\bf{n}})$ contributions and $M_\ell$ which
describes their relative contribution to each multipole moment.  Given the
discussion above, we seek a model where the intrinsic anisotropy is maximally
modulated for the quadrupole and minimally modulated beyond the octopole.  Note
that the model is constructed so that in the absence of modulation ($w_2=0$)
$C_\ell = C_\ell^{\rm fid}$ for $f=1$.  Hence for simplicity we set $M_2=1$ and
$M_{\ell \ge 4}=0$.  The proof of principle model is then defined by three
parameters $w_2$, $M_3$, and $f$.
 
 \subsection{Power Deficits and Alignments}
 \label{sec:alignments}
 
 We now assess whether the WMAP data favor such a multiplicative
 modulation.  We again begin by considering the likelihood
\begin{eqnarray} 
{\cal L}({\bf t} | {\bf C}) &=&
{ 1 \over (2\pi)^{N/2} \sqrt{{\rm det {\bf C} }}}
\exp \left[ -{1\over 2} {\bf t}^\dagger {\bf C}^{-1} {\bf t} \right]
\nonumber\\
&\equiv& \exp [-\chi_{\rm eff}^2/2]
\end{eqnarray}
where ${\bf t}$ is an $N$ element vector of the WMAP TOH multipole moments
$t_{\ell m}$ in the dipole direction.  ${\bf C}$ is the sum of the signal
covariance matrix in \eq{eqn:covmatrix} and a diagonal noise covariance given
by $N_\ell$ in Tab.~\ref{tab:tlms}.  The signal covariance matrix is
parameterized by $w_2$, $M_3$ and $f$ [see Eqs.~(\ref{eqn:covmatrix})
and (\ref{eqn:toymodel})]. It is important here to include $N_\ell$
since some of the multipole moments are consistent with noise and the model has
the ability to lower the predicted power in certain multipoles to zero.  On the
other hand an accurate quantification of noise is not crucial.

In Fig.~\ref{fig:chi2eff_dip}, we show the improvement in $\Delta\chi_{\rm
eff}^2$ relative to the fiducial model as a function of $(w_2,M_3)$ for the
value of $f$ that maximizes the likelihood.  The maximum likelihood value at
$w_2=-5.0$ and $M_3=0.82$ has a $\Delta \chi^2_{\rm eff}=-16$.  Even more
striking, most of the gain comes through $w_2$.  The best unmodulated or
statistically isotropic model (restricted to $w_2=0$) has only an improvement
of $\Delta \chi^2_{\rm eff}\approx-5$.  Note that in this case maximization
over $(f,M_3)$ corresponds to the a maximization over $(C_2,C_3)$.  The maximum
at $w_2 = -5.0$ is driven by the most anomalously low multipole moment in the
dipole direction $\ell=2$, $m=0$ as shown in Fig.~\ref{fig:magicw2}.  Note
however that a wide range of values surrounding the maximum would also improve
the fit.

To test the quadrupole-octopole alignment,  we take the normalized angular momentum \citep{deOliveira2004} as generalized by Ref. \citep{Copi2005}
\begin{equation}
 \hat L^2_\ell \equiv { \sum_{m=-\ell}^\ell m^2 \vert a_{\ell m}\vert^2 \over 
\ell^2 \sum_{m=-\ell}^\ell \left| \alm \right|^2  } \, . \slabel{eqn:angmomn}
\end{equation}
The statistic
\begin{equation} 
\hat L^2\equiv {1\over 2} \left ( \hat L^2_2 +
 \hat L^2_3 \right ) \,
\end{equation}
maximized over direction of the preferred axis captures both the alignment of
the quadrupole and octopole and the planar nature of the octopole.  Using
\eqs{eqn:covrot}{eqn:angmomn} and maximizing over direction, we find the best
fit axis to be \beq l=247 \fdg 3,\quad b=59 \fdg 7 \slabel{eqn:axisofevil} \eeq
in galactic coordinates.  This is close to but differs from the direction of
the dipole $(l=263 \fdg 85\pm0 \fdg 1, b=48 \fdg 25\pm0 \fdg 04)$ \cite{Bennett2003}.
For WMAP $\hat L^{2}_{\rm WMAP} = 0.96$.
 
 To assess whether the multiplicative modulation improves the probability of
 alignments, we draw Monte Carlo realization of the models.  With the fiducial
 model only $0.22\%$ of $10^4$ realizations had a $ \hat L^2 >  \hat
 L_{\rm WMAP}^2$ in any direction.  For the multiplicative modulation models, the
 fraction of $10^3$ realizations is shown in Fig.~\ref{fig:alignments_dip}.
 Here $f$ is fixed to minimize $\Delta\chi_{\rm eff}^{2}$ for each $(w_2,M_3)$
 in the dipole direction
 
 This fraction is maximized at $w_2=-7$ and $M_3 =1$ at 45\%, an increase of a
 factor of $\sim 200$ over the fiducial model.  From Fig.~\ref{fig:magicw2}, we
 see that $w_2 = -7$ achieves the best balance of reducing the power in all
 moments of the quadrupole and octopole except $m=\pm \ell$.  Hence such a
 model would be expected to produce the highest normalized angular momentum.
In Fig.~\ref{fig:maps} we show an example of one of these realizations 
(with $\hat L^2 =0.99$) that illustrates
how the modulation takes an intrinsically isotropic sky into one that exhibits
the alignments of WMAP.

\begin{figure}
\epsfig{file=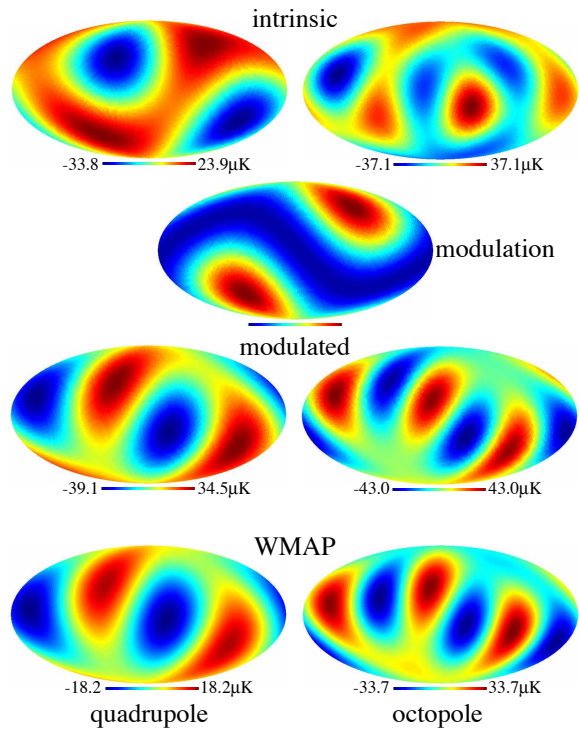, width=3in}
\caption{\footnotesize \slabel{fig:maps} A realization of the multiplicative
model where the quadrupole (left column) and octopole (right column) 
exhibit an alignment similar to WMAP.
First row: intrinsic (unmodulated) sky from a Gaussian random isotropic
realization.  Second row (single column): the quadrupolar modulation
$ \propto -[1-7 Y_{20}(\hat{\bf n})]$ in the dipole direction.
Third row: the modulated sky of the observed CMB (with $M_{3}=1$).  
Fourth row: WMAP full-sky quadrupole and octopole.  }
\end{figure} 

Moreover, we have verified that models that solve the alignment problems do so
irrespective of which test is used to quantify  them. For example, the aforementioned
model with $w_2=-7$ and $M_3=1$ makes the mutual closeness of the quadrupole and
octopole area vectors \citep{Schwarz2004, Copi2005} unlikely at 37\% level, compared
to 0.12\% level without modulation. 
 
The alignment statistic also favors high values of $M_3$ since $M_2=M_3=1$
implies that all contributions to the quadrupole and octopole are modulated.
The $\chi^2_{\rm eff}$ statistic slightly disfavors very high values since in
the dipole direction there remains a substantial amount of power in $\ell=3$,
$m=\pm 1, \pm 2$ which is better fit with an uncorrelated isotropic component
(see Tab.~\ref{tab:tlms}).  Nonetheless there is a range of models around
$w_2=-7$ and $M_3=0.95$ that both improve $\chi^2_{\rm eff}$ and produce a
fraction of more extremely aligned skies of $>10\%$.
 
 \begin{figure}
\epsfig{file=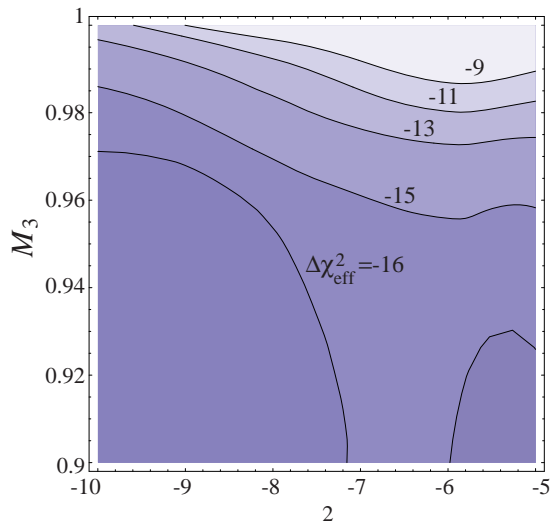, width=3in}
\caption{\footnotesize \slabel{fig:chi2eff_lmax} Same as
Fig.~\ref{fig:chi2eff_dip} except replacing the dipole as the preferred axis
with that of $ \hat L_{\rm max}^2$.  By maximizing the planarity of the
quadrupole and octopole $|m| < \ell$ modes are minimized allowing for smaller
$w_{2}$ and higher $M_{3}$. }
\end{figure} 

\begin{figure}
\epsfig{file=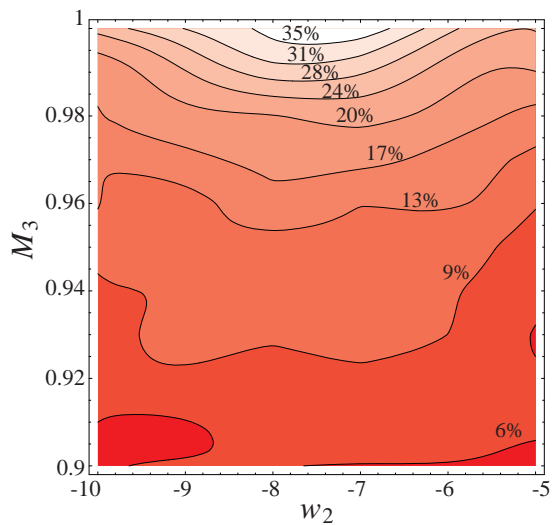, width=3in}
\caption{\footnotesize\slabel{fig:alignments_lmax} Same as in
 Fig.~\ref{fig:alignments_dip} except with $f$ given by the minimum $\Delta
 \chi^2_{\rm eff}$ in the $ \hat L_{\rm max}^2$ direction.  Alignment
 requirements remain largely unchanged from the dipole direction but maximum
 percentages are more compatible with substantial improvements in $\Delta\chi^{2}_{\rm eff}$
 from Fig.~\ref{fig:chi2eff_lmax}.}
\end{figure} 

In fact, even better models can be found if so desired.  Choosing the preferred
axis to be exactly the dipole for parameter fitting is not necessary.  In fact
in the $ \hat L_{\rm max}^2$ direction of WMAP (see \eq{eqn:axisofevil})
the planarity of the observed quadrupole and octopole is maximized and hence
has particularly low $m=\pm 1$, $m =\pm 2$ components (see
Tab.~\ref{tab:tlms}).  Fitting parameters of the model to the data with this as
the preferred axis allows a wider range of solutions (see
Fig.~\ref{fig:chi2eff_lmax}).  Now models with more negative $w_2$ are favored to
reduce the power in these components and higher values of $M_3$ are allowed.
With the maximum likelihood $f$ values with this preferred axis, $w_2=-7$ and
$M_3 \approx 1$ still produces more extreme alignments in $\sim 45\%$ of the
Monte Carlos.  Moreover models exist with $\chi^2_{\rm eff}$ improvements of
more than $-10$ which produce alignments more extreme than WMAP in $\simgt
30\%$ of the Monte Carlos (see Fig.~\ref{fig:alignments_lmax}).

The change in $ \Delta \chi^2_{\rm eff}$ between the dipole and $\hat L_{\rm max}^2$ directions
and more generally its sensitivity to direction
indicates that one should be careful in interpreting this improvement probabilistically.
A $\Delta \chi^2_{\rm eff}\approx -16$ likelihood improvement
naively suggests a ``4-$\sigma$'' result.  However since the choice of the $\hat L_{\rm max}^2$ direction 
maximizes the planarity of the quadrupole and octopole, there will be some preference
for modulation even for realizations of a statistically isotropic sky.

 To address this issue, we fit the model parameters in the $10^4$ 
realizations of the isotropic fiducial model in the $\hat L_{\rm max}^2$ direction of
each run.  We find that the WMAP measured sky shows a greater
improvement in the likelihood than 98.8\% of the Gaussian random, isotropic
skies indicating that the multiplicative model is indeed strongly
preferred by the WMAP observations.
 
Since this exercise is simply a proof of principle with somewhat ad hoc
parameters, we do not pursue finding the best model parameters further.  The
important conclusion is that models exist where a multiplicative modulation of
large angle anisotropies in a quadrupolar pattern can make the WMAP alignments
a likely occurrence.

\section{Discussion}
\label{sec:discussion}

We have introduced a general class of models that break statistical isotropy in
CMB temperature fluctuations spontaneously.    The key elements are a non-linear
response in  the CMB to  long wavelength fluctuations in a mediating field. 
Long wavelength fluctuations pick a direction through their local gradient and
a non-linear response allows this direction to be imprinted in a range of
multipole moments.

We have examined two subclasses of models based on whether isotropy is
broken through an additive contribution to the anisotropy or a multiplicative
one.  The two classes have analogues in the instrumental domain and are
related to saturation in detectors and an angular drift in calibration. 

As a cosmological example of the additive class, we considered gradients in a
quintessence field that are non-linearly mapped through that field's potential
into dark energy density perturbations.  The CMB then responds linearly to
these density perturbations through the integrated Sachs-Wolfe effect.  
The existence of such contributions allows for the underlying CMB to be
isotropic.  Nonetheless, additive contributions cannot be said to solve the
alignment problem in WMAP.  Since the anomalies in the quadrupole and octopole
are related to a deficit of power near the dipole direction, an additive
correction can only explain the observations if the intrinsic fluctuations
cancel them in both.  This is statistically less likely than having an aligned
deficit of power by chance without isotropy breaking.

This fundamental problem exists for all explanations of the alignments which
seek to restore power to the CMB including those involving foregrounds or other
uncorrelated templates.  Additionally, an additive contribution based on an
azimuthally symmetric modulation of a gradient field can only affect $m=0$
multipoles around the preferred axis.  The only remaining possibility is that
the additive contributions themselves account for most of the power in the
quadrupole and octopole.  All of the moments of the intrinsic quadrupole and
octopole would then be
even lower than those observed, exacerbating the power deficit problem.

These difficulties can be overcome by a model which breaks isotropy
multiplicatively.  Here we have demonstrated a proof of principle that the
alignments and power distribution of WMAP can be made likely with such a
mechanism.  By modulating the intrinsic quadrupole and octopole in a
quadrupolar fashion, the likelihood of the WMAP data can be increased by a
factor of $e^{16/2}$ and separately, the probability of obtaining a sky with a
higher angular momentum statistic is increased 
by $200$ to $\sim 45\%$.  
The angular momentum 
statistic is a quantification of both the planarity of the octopole and its
alignment with the quadrupole.  
Note that the likelihood increase must be considered in the context of
the directional maximization of the angular momentum.   The fiducial isotropic model
would falsely produce greater than $e^{16/2}$ improvements 
in $\sim 1\%$ of realizations.

In addition, the model predicts correlated
power excess power in $\ell=3$, $m=3$ and $\ell=5$, $m=3$ and more generally
the $\Delta \ell=2$ parity pattern of anomalies in the data \cite{Land2005c}.
There is a range of parameters that jointly increase the likelihood by
$e^{10/2}$ and the angular momentum probability to $\sim 30\%$.

There are also classes of anomalies that are not explained by this version of
spontaneous isotropy breaking.   Solar system related alignments
 \citep{Schwarz2004,Copi2005} and the asymmetry in
power between the northern and southern ecliptic ecliptic hemisphere
\citep{Eriksen_asym} both fall into this category.  While isotropy breaking
involving a dipolar modulation \cite{Pruetal04} along the ecliptic could potentially explain
the latter feature, it cannot simultaneously explain anomalies along the dipole
direction.  Furthermore, there is no fundamental link between the dipole of our
peculiar motion and the preferred axis for symmetry breaking.  That coincidence
remains in the multiplicative model.

We emphasize that this proof of principle is not a physical model and
challenges will have to be overcome in constructing one.  The level of
multiplicative modulation preferred is fairly extreme and cannot be considered
a perturbation.  It must affect essentially all of the quadrupole and octopole
in spite of the fact that in the fiducial cosmology there are two physically
distinct sources to these fluctuations: the Sachs-Wolfe and ISW effects.  Thus
a physical solution will have to modify the power spectrum of these
contributions as well.  On the other hand the modulation must not continue to
intrinsic fluctuations in multipoles much higher than the octopole.  Thus two
distinct sources of anisotropy are required.

Constructing a physical mechanism that satisfies all of these requirements will
clearly be challenging and is beyond the scope of this paper. Likewise the
significance of the alignments and hence the model improvements may be degraded
by foreground and systematic contamination not accounted for in the all-sky
cleaned maps, at the expense of exacerbating the power deficit
problems.  Nevertheless, this proof of principle model can serve as a guide
both to building a physical model and to searching for alternate explanations
of the anomalies.

\smallskip{\it Acknowledgments:} CG was supported by the KICP under NSF
PHY-0114422; WH by the DOE and the Packard Foundation; DH by the NSF Astronomy
and Astrophysics Postdoctoral Fellowship under Grant No.~0401066; TC by NSF
OPP-0130612. We have benefited from using the publicly available
Healpix package~\cite{healpix}. We thank Stephan Meyer, Hiranya Peiris
and Bruce 
Winstein for useful discussions and Craig Copi for providing a Wigner rotation
matrix code.  WH thanks the WMAP team for supplying the ``beach ball"
representation of the data.


\end{document}